\documentclass[12pt]{article}
\usepackage{amssymb}
\usepackage{epsfig}

\catcode`\@=11
\def\numberbysection{\@addtoreset{equation}{section}
        \def\theequation{\thesection.\arabic{equation}}}
\def\beq{\begin{equation}}
\def\eeq{\end{equation}}
\numberbysection
\begin{document}
\begin{titlepage}
\begin{center}
\hfill DFF  1/10/98 \\
\vskip 1.in
{\Large \bf Integrability of the $N$-body problem in ($2+1$)-$AdS$ gravity}
\vskip 0.5in
P. Valtancoli
\\[.2in]
{\em Dipartimento di Fisica dell' Universita', Firenze \\
and INFN, Sezione di Firenze (Italy)}
\end{center}
\vskip .5in
\begin{abstract}
We derive a first order formalism for solving the scattering of point
sources in $(2+1)$ gravity with negative cosmological constant. We
show that their physical motion can be mapped, with a polydromic
coordinate transformation, to a trivial motion, in such a way that
the point sources move as time-like geodesics 
( in the case of particles ) or as space-like geodesics ( in the case of
$BTZ$ black holes ) of a three-dimensional hypersurface immersed in a
four-dimensional Minkowskian space-time, and that the
two-body dynamics is solved by two invariant masses, whose difference
is simply related to the total angular momentum of the system. 
\end{abstract}
\medskip
\end{titlepage}
\pagenumbering{arabic}
\section{Introduction}

   In this article we would like to study the problem of treating the 
interaction between particles and $AdS$ gravity, generalizing our previous
investigation in $(2+1)$ gravity and its supergravity extension, which has 
also been considered by many other authors
\cite{a1}-\cite{a2}-\cite{a3}-\cite{a21}-\cite{a13}-\cite{a14}-\cite{a15}-
\cite{a11}-\cite{a12}-\cite{00}. The problem from a
dynamical point of view is quite difficult, because the introduction
of an explicit scale parameter ( the cosmological constant ) in the  
theory induces a static Newtonian force between point sources, whose absence
was one of the main features that led to the solution of the two body problem 
in $(2+1)$ gravity. Nevertheless one has to expect surprising
simplifications in the dynamics, since the Lagrangian of $AdS$ gravity
splits in two Chern-Simons theories, as Witten has shown in ref. \cite{a2}.

Our aim is to clarify in what sense the interaction given by 
$AdS$ gravity can be considered integrable. First of all, we have asked ourself
if the general scheme for solving $(2+1)$ gravity, described in 
\cite{a13}-\cite{a14},  can be generalized 
in a continuous way to the case of negative cosmological constant (
hereafter indicated by its modulus $\Lambda$ ). Our proposal is that
it is still
possible to reduce the solution of the equations of motion to
the knowledge of a polydromic mapping $X^A = X^A(x^a)$, and that it exists
a generalized Minkowskian system where the motion of the interacting particles
is almost free, apart from the property that they are constrained to move on a
hypersurface, instead of moving freely in a plane. The price to pay is
that we have to introduce this polydromic mapping as an immersion in
four dimensions with a quadratic constraint between the coordinates:

\beq ds^2 = dX^A dX^B \eta_{AB} \ \ \ \ \ \ \
X^A X^B \eta_{AB} = {( X^0 )}^2 + {( X^1 )}^2 - {( X^3 )}^2 - {( X^4 )}^2
\ = \ \frac{1}{\Lambda}. \label{a0}
\eeq

Due to this constraint, the general allowed polydromy is reduced to
the Lorenz subgroup $SO(2,2)$
\beq X^A \rightarrow L^A_B \ X^B \label{a1} \eeq 
which has the same number of generators of $ISO(2,1)$.

This procedure is not new, because it has already been discussed in several
contexts, for example to introduce the $BTZ$ black hole in $(2+1)$
dimensions \cite{a4}-\cite{a5}.
Our proposal is to unify several partial results in a unitary program,
analogously to what has been successfully done in the case of 
$(2+1)$ gravity with point sources \cite{a14}. 
This investigation should shed also light on the scattering of
$BTZ$ black holes, which can contain some important features in common 
with the four dimensional case, and could open new issues for the
quantum case \cite{a8}.

\section{Examples of polydromic mappings in $AdS$-gravity}

In general, the conical cuts of the particles in ($2+1$) gravity ,
firstly defined in \cite{a1}, must
be substituted by similar cuts related to the group $SO(2,2)$ instead
of $ISO(2,1)$, and therefore the translation of the center of rotation
must also be represented with a boost of $SO(2,2)$.
While the general case will be postponed to the next section, we are going to 
show explicit examples of polydromic mappings for static bodies, and in that
case the polydromy can still be reduced to a rotation.

Let us start with the metric of a single body, which is given in the radial 
gauge as :

\beq ds^2 = ( {(1-\mu)}^2 + \Lambda r^2 ) dt^2 - \frac{dr^2 }{ 
{(1-\mu)}^2 + \Lambda r^2 } - r^2 d\theta^2 . \label{a2} \eeq

This can be expressed as a polydromic mapping

\begin{eqnarray}
X^t \ & = & \ X^0 + i X^1 \ = \ \frac{1}{\sqrt{\Lambda}} \sqrt{ 1 + 
\frac{\Lambda r^2 }{{(1-\mu)}^2} } e^{i \sqrt{\Lambda} (1-\mu) t} 
\nonumber \\
X^z \ & = & \ X^2 + i X^3  \ = \ \frac{r e^{i(1-\mu)\theta} }{ (1-\mu) }
\label{a3} \end{eqnarray}
of the flat metric (\ref{a0}) that respects the quadratic constraint, where the
polydromy is a pure rotation

\beq X^z \rightarrow e^{ -2 \pi i \mu} X^z \label{a4} . \eeq

The equation of motion for the particle source 
in the $X^a$ coordinates is described by the parametric equation 
\beq X^2 = X^3 = 0 \ \ \ \ X^0 = \frac{1}{\sqrt{\Lambda}} cos ( \sqrt{\Lambda}
s ) \ \ \ X^1 = \frac{1}{\sqrt{\Lambda}} sin ( \sqrt{\Lambda} s ) ,
\label{a5} \eeq
and by definition it corresponds to a time-like geodesics:
\beq {\left( \frac{dX^0}{ds} \right)}^2 + 
 {\left( \frac{dX^1}{ds} \right)}^2 = 1 \ > 0 \ . 
\label{a6} \eeq

As a second example let us consider the metric of a spinning particle

\beq ds^2 = dt^2 ( {(1 - \mu )}^2 + \Lambda r^2 ) + 
J dt d\theta - r^2 d\theta^2 - 
\frac{ dr^2 }{ {(1 - \mu )}^2 + \Lambda r^2 + \frac{J^2}{4r^2}} 
\label{a7} \eeq
which can be obtained by the following mapping

\begin{eqnarray}
X^t & = & \frac{1}{\sqrt{\Lambda}} \left( \sqrt{ \frac{ r^2 + r^2_{+} }{
 r^2_{+} - r^2_{-} } } \right) e^{i ( \Lambda r_{+} t + \sqrt{\Lambda} r_{-}
\theta )} \nonumber \\
X^z & = & \frac{1}{\sqrt{\Lambda}} \left( \sqrt{ \frac{ r^2 + r^2_{-} }{
 r^2_{+} - r^2_{-} } } \right) e^{i ( \Lambda r_{-} t + \sqrt{\Lambda} r_{+}
\theta )} , \label{a8}
\end{eqnarray}
where the two roots $r_{\pm}$ are defined as
\beq r^2_{+} + r^2_{-}  =  \frac{{(1 - \mu )}^2}{\Lambda}
\ \ \ \ \ \  r^2_{+} r^2_{-} = \frac{J^2}{4\Lambda} .
\label{a9} \eeq

This metric realizes the elliptic monodromy

\begin{eqnarray}
X^t & \rightarrow & e^{2i \pi \sqrt{\Lambda} r_{-}} \ X^t \nonumber \\ 
X^z & \rightarrow & e^{2i \pi \sqrt{\Lambda} r_{+}} \ X^z . 
\label{a10} \end{eqnarray}

We characterize with the word elliptic monodromy every monodromy
which can be reduced by a similitude transformation to a pure
rotation. In this way we distinguish it from a $BTZ$ black hole, which 
we can call an hyperbolic monodromy, since it can reduced by a similitude 
transformation to a pure boost. It is useful then to remember how the
$BTZ$ black hole has been introduced in refs. \cite{a4}-\cite{a5}.
In the general case of the spinning black hole metric, where the line element
is given by

\beq ds^2 = ( \Lambda r^2 - M^2 ) dt^2 + J dt d \theta - r^2 d \theta^2 - 
\frac{dr^2}{\Lambda r^2 - M^2 + \frac{J^2}{4r^2}} ,
\label{a11} \eeq
the $X^A$-mapping can be written in terms of two radii $r_{\pm}$:
\beq r_{\pm} = \frac{M}{\sqrt{2\Lambda}} \sqrt{ 1
\pm \sqrt{1 - \frac{\Lambda J^2}{M^4}} } 
\label{a12} \eeq
as follows:

\begin{eqnarray}
r > r_{+} \ \ \  X_0 \pm X_2 & = & \frac{1}{\sqrt{\Lambda}} 
\sqrt{\frac{r^2 - r^2_{-}}{r^2_{+}- r^2_{-}}} e^{\pm \Theta (t,
\theta)} \nonumber \\
 X_1 \pm X_3 & = & \pm \frac{1}{\sqrt{\Lambda}} 
\sqrt{\frac{r^2 - r^2_{+}}{r^2_{+}- r^2_{-}}} e^{\pm T (t,
\theta)} \nonumber \\
r_{-} < r < r_{+} \ \ \  X_0 \pm X_2 & = & \frac{1}{\sqrt{\Lambda}} 
\sqrt{\frac{r^2 - r^2_{-}}{r^2_{+}- r^2_{-}}} e^{\pm \Theta (t,
\theta)} \nonumber \\
 X_1 \pm X_3 & = & \frac{1}{\sqrt{\Lambda}} 
\sqrt{\frac{r^2_{+}- r^2}{r^2_{+}- r^2_{-}}} e^{\pm T (t,
\theta)} \nonumber \\
0 < r < r_{-} \ \ \  X_0 \pm X_2 & = & \pm \frac{1}{\sqrt{\Lambda}} 
\sqrt{\frac{r^2_{-} - r^2}{r^2_{+}- r^2_{-}}} e^{\pm \Theta (t,
\theta)} \nonumber \\
 X_1 \pm X_3 & = &  \frac{1}{\sqrt{\Lambda}} 
\sqrt{\frac{r^2_{+} - r^2}{r^2_{+}- r^2_{-}}} e^{\pm T (t,
\theta)} , \label{a13} \end{eqnarray}
where  the following functions are defined, as in ref. \cite{a5}:
\begin{eqnarray} 
T ( t, \theta ) & = & \Lambda r_{+} t - \sqrt{\Lambda} r_{-} \theta 
= \frac{1}{2} ln \left( \frac{ X^1 + X^3 }{ X^1 - X^3 } ( \theta 
( r_{+} - r ) - \theta ( r - r_{+} ) ) \right) 
\nonumber \\
\Theta ( t, \theta ) & = & \sqrt{\Lambda} r_{+} \theta - \Lambda r_{-} t 
= \frac{1}{2} ln \left( \frac{ X^0 + X^2 }{ X^0 - X^2 } ( \theta 
( r - r_{-} ) - \theta ( r_{-} - r ) ) \right) .
\label{a14} \end{eqnarray}

This metric realizes the hyperbolic monodromy:
\begin{eqnarray}
 X^0 \pm X^2 & \rightarrow & e^{\pm 2 \pi \sqrt{\Lambda} r_{+} } 
( X^0 \pm X^2 ) \nonumber \\
 X^1 \pm X^3 & \rightarrow & e^{\mp 2 \pi \sqrt{\Lambda} r_{-} }
( X^1 \pm X^3 ) .
\label{a15} \end{eqnarray}

In the simpler case $J=0$, the $X^A$-mapping reduces to

\beq X^0 \pm X^2 = \frac{r}{M} e^{\pm \theta M } \ \ \ \ 
X^1 \pm X^3 = \pm \frac{1}{\sqrt{\Lambda}} \ 
\left( \sqrt{ \frac{\Lambda r^2}{M^2} -1 } \right) 
e^{\pm \sqrt{\Lambda} M t } \label{a16} \eeq
in the region external to the black hole ($ r > \frac{M}{\sqrt{\Lambda}}$),
while
\beq X^0 \pm X^2 = \frac{r}{M} e^{\pm \theta M} \ \ \ \ \
X^1 \pm X^3 = \frac{1}{\sqrt{\Lambda}} \left( 
\sqrt{ 1 - \frac{\Lambda r^2}{M^2}} \right) \ 
e^{\pm \sqrt{\Lambda} M t} \label{a17} \eeq
in the internal region.

Here the equation of motion for the source in the
$X^a$-coordinates is described by the parametric equation

\beq X^0 = X^2 = 0 \ \ \ \ X^1 = \frac{1}{\sqrt{\Lambda}} ch ( \sqrt{\Lambda}
s ) \ \ X^3 = \frac{1}{\sqrt{\Lambda}} sh ( \sqrt{\Lambda} s ) \label{18}\eeq
and by definition it corresponds to a space-like geodesics:  
\beq {\left( \frac{dX^1}{ds} \right)}^2 -  {\left( \frac{dX^3}{ds} \right)}^2
= -1 < 0  . \label{a19} \eeq

To verify the notion that point sources move freely in the Minkowskian
coordinate system $X^A$ as geodesics of the (\ref{a0}) hypersurface, we 
are going to derive again a known result, the determination of the geodesics 
around the spinning black hole \cite{a6}, making a bridge between our
first order formalism, which is different from what physicists call
first order or dreibein formalism ( see for example ref. \cite{a7}),
and the standard method based on integrating the geodesic equations. 

From the $X^A$-mapping (\ref{a13})-(\ref{a14}) it is easy to read 
directly the solution to the geodesic equations:
\begin{eqnarray}
r^2 (\tau) & = & r^2_{-} + \Lambda ( r^2_{+} - r^2_{-} ) 
[ ( X^0 )^2 - ( X^2 )^2 ] (\tau) \nonumber \\
\theta (\tau) & = & \frac{ r_{+} \Theta (\tau) + r_{-} T (\tau) 
}{ \sqrt{\Lambda} ( r^2_{+} - r^2_{-} ) }
\nonumber \\
t (\tau) & = &  \frac{ r_{+} T(\tau) + r_{-} \Theta (\tau) 
}{ \Lambda ( r^2_{+} - r^2_{-} ) } .
\label{a20} \end{eqnarray} 
We simply need to complete eq.(\ref{a20}) with the general
parameterization describing the motion of a test body $X^A = X^A (\tau)$
on the $X^A X_A = 1 /\Lambda $ hypersurface and satisfying 
\beq \ddot{X}^A (\tau) = - \Lambda m X^A (\tau) ,
\label{a21} \eeq
where $m = \dot{X}^A \dot{X}_A  = 1, 0, -1$ for time-like, null and 
space-like geodesics.

A general parameterization of a geodesic is given by :
\begin{eqnarray}
X^A (\tau) & = & c_0^A cos( \sqrt{\Lambda} \tau ) + c_1^A 
sin (\sqrt{\Lambda} \tau ) \ \ \ \ \ \ \ {\rm if} \ m = 1 \nonumber \\
    & = & c_0^A  + c_1^A \sqrt{\Lambda} \tau  
\ \ \ \ \ \ \ \ \ \ \ \ \ \ \ \ \ \ \ \
\ \ \ \ \ \ \ \ {\rm if} \ m = 0 \nonumber \\
& = & c_0^A cosh ( \sqrt{\Lambda} \tau ) + c_1^A 
sinh (\sqrt{\Lambda} \tau ) \ \ \ \ \ \ \ {\rm if} \ m = - 1 ,
\label{a22} \end{eqnarray}
where the vectors $( c^A_0 , c^A_1)$, constants of
motion, have to satisfy :
\begin{eqnarray}
c_0^A c_{0A} & = & \frac{1}{\Lambda} \ \ \ \ \ \ \ 
c_0^A c_{1A} = 0 \ \ \ \ \ \ \ c_1^A c_{1A} =   \frac{m}{\Lambda} .
\label{a23} \end{eqnarray}

By using the rescaled variables 
\begin{eqnarray}
r' & = & \frac{\sqrt{\Lambda}}{M} r \ \ \ \ \ 
r'_{\pm} = \frac{\sqrt{\Lambda}}{M} r_{\pm} \ \ \ \ \ 
\theta' = M \theta \ \ \ \ \ 
t' = \sqrt{\Lambda } M  t \nonumber \\
\tau' & = & \sqrt{\Lambda} \tau \ \ \ \ \ 
E' = \frac{E}{M} \ \ \ \ \ 
L' = \frac{\sqrt{\Lambda}}{M} L \ \ \ \ \
J' = \frac{\sqrt{\Lambda}}{M^2} J  ,
\label{a24} \end{eqnarray}
neglecting the primes from now on, we can compare these total
integrals ( eqs. (\ref{a20}) and (\ref{a22}) )
with the first-integrals found in ref. \cite{a6}:
\begin{eqnarray}
{\left(\frac{dr}{d\tau}\right)}^2 
& = & - m ( r^2 - 1 + \frac{J^2}{4r^2}) 
+ E^2 - L^2 +  \frac{L^2 - J E L }{r^2} 
\nonumber \\
\frac{d\theta}{d\tau} & = & \frac{(r^2 - 1) L + \frac{1}{2} J E }{
( r^2 - r^2_{+} ) ( r^2 - r^2_{-} )} \nonumber \\
\frac{dt}{d\tau} & = & \frac{E r^2 - \frac{1}{2} J L }{(r^2 - r^2_{+})
( r^2 - r^2_{-} )} .
\label{a25} \end{eqnarray}
From eq. (\ref{a21}) we notice that each of the following six 
combinations $P^{AB} \ = \ \Lambda
( \dot{X}^A X^B - \dot{X}^B X^A ) $ ( dot means $\frac{d}{d\tau'}$
which has been relabelled $\frac{d}{d\tau}$ after eq. (\ref{a24}) )
represents a constant of motion along a geodesics. Between them only
two are globally defined, i.e. they are invariant under the intrinsic 
polydromies of the $X^A$-coordinates, and correspond to 
\begin{eqnarray}
P^{20} & = & \Lambda ( \dot{X}^2 X^0 - \dot{X}^0 X^2 ) = 
\Lambda ( c^0_0 c^2_1 - c^0_1 c^2_0 ) \ = \ \frac{1}{\sqrt{ 1 - J^2}} \
\left( L r_{+} - \frac{1}{2} \frac{J E}{r_{+}} \right) 
\nonumber \\
P^{13} & = &  \Lambda ( \dot{X}^1 X^3 -\dot{X}^3 X^1 ) =  
\Lambda ( c^3_0 c^1_1 - c^3_1 c^1_0 ) \ = \ 
\frac{1}{\sqrt{ 1 - J^2}} \
\left( \frac{1}{2} \frac{J E}{r_{-}} - L r_{-} \right) , 
\label{a26} \end{eqnarray} 
where $L$ is the angular momentum and $E$ is the energy as defined  in 
\cite{a6}.

By developing eqs. (\ref{a20}) and (\ref{a22}) one arrives at the
following complete integrals\footnote{In the rescaled variables 
$r_{\pm} = \frac{1}{\sqrt{2}} \sqrt{ 1 \pm \sqrt{ 1 - J^2 }}$ . }:

\begin{eqnarray}
r^2 (\tau) & = & \left\{ 
\begin{array}{ll}
\frac{1}{2} [ A + C \ sin 2 ( \tau - \tau_0 ) ] &
\mbox{if \ $m = 1$} \\
A {( \tau - \tau_0 )}^2 - \frac{B}{A} &  \mbox{if \ $m = 0$ \ and $A
\neq 0$ } \\
2\sqrt{B} ( \tau - \tau_0 )  &  \mbox{if \ $m = 0$, \   $A
= 0$ \ and $B \neq 0$ } \\
{\rm cost. } &  \mbox{if \ $m = 0$, \ $A = B = 0$ } \\
\frac{1}{2} [ - A + C \ cosh 2 (  \tau - \tau_0 ) ] &
\mbox{if \ $m = -1$} 
\end{array} \right. \nonumber \\ 
A & = &  E^2 - L^2 + m \ \ \ \ \ \ \ \ B = L^2 - J E L - \frac{1}{4} m J^2  
\nonumber \\
C & = & \sqrt{ A^2 + 4 m B } = \sqrt{ {F_{\pm}}^2 + 4 m G^2_{\pm} }
\nonumber \\
F_{\pm} & = & E^2 - L^2 + m ( 1 - 2 r^2_{\pm} ) \ \ \ \ \ \ 
G_{\pm} = r_{\pm} \left( E - \frac{JL}{2r^2_{\pm}} \right) 
\label{a27} \end{eqnarray}
\begin{eqnarray}
\theta(\tau) & = & \frac{ r_{+} \Theta (\tau) + r_{-} T(\tau ) 
}{ r^2_{+} - r^2_{-} } \nonumber\\
t(\tau) & = & \frac{ r_{-} \Theta (\tau) + r_{+} T(\tau ) 
}{ r^2_{+} - r^2_{-} }  \nonumber\\
\Theta (\tau) & = & \frac{1}{2} ln \left[ 
\frac{ ( c^0_0 + c^2_0 ) f'_m(\tau)  
+ ( c^0_1 + c^2_1 ) f_m(\tau)}{
( c^0_0 - c^2_0 )  f'_m(\tau)
+ ( c^0_1 - c^2_1 )  f_m(\tau)} 
( \theta( r - r_{-} ) - \theta( r_{-} - r ) )
\right]  = \ \ \ \ \nonumber \\
& = & - \frac{1}{2} ln \left| \frac{ 
( r^2 - r^2_{-} ) }{
2 G_{-}^2 +  F_{-} ( r^2 - r^2_{-} )  + 2 G_{-} \sqrt{
G_{-}^2 + F_{-} ( r^2-r^2_{-} ) - m {( r^2- r^2_{-} )}^2} } \right| + \Theta_0
\nonumber \\
T(\tau) & = & \frac{1}{2} ln \left[
\frac{ ( c^1_0 + c^3_0 )  f'_m(\tau)
+ ( c^1_1 + c^3_1 ) f_m(\tau)}{
( c^1_0 - c^3_0 )  f'_m(\tau) 
+ ( c^1_1 - c^3_1 ) f_m(\tau)} 
( \theta ( r_{+} - r(\tau)) - \theta ( r(\tau) - r_{+} ) ) \right] = 
\nonumber \\
& = & \frac{1}{2} ln \left|\frac{  
( r^2 - r^2_{+} ) }{
2 G_{+}^2 +  F_{+} ( r^2 - r^2_{+} )  + 2 G_{+} \sqrt{
G_{+}^2 + F_{+} ( r^2-r^2_{+} ) - m {( r^2- r^2_{+} )}^2}
} \right| + T_0 , \nonumber \\
& & \label{a28} 
\end{eqnarray}
where $f_m(\tau) \ = \ ( sin (\tau) , \tau, sh (\tau) )$ \ \ if 
$m = (1, 0, -1)$, and  use has been made of the formulas given in the Appendix.

Therefore this method reproduces all the previous results \cite{a6}, and, 
in our opinion, is more flexible to be generalized to the $N$-body
problem, discussed also in ref. \cite{a9}.

\section{N body -problem}

Now we are going to define the integrability of a system of
$N$ point sources interacting with $AdS$ gravity, 
as a completely non-interacting system in the
Minkowskian $X^A$-coordinates, apart from the fact that their motion has to
respect the constraint $ X^A X_A = \frac{1}{\Lambda} $. We have just
shown that, at least for the one body case, the effect of a point
source, particle or $BTZ$ black hole, is to eliminate a portion of the 
hypersurface, while a test particle moves as a geodesic of it. In the
interacting case we can suppose that each moving point source carries 
its deficit angle and that it doesn't scatter until it reaches the
extremity of the cut of another one. Therefore
we expect that their scattering is again topological, i.e. it can be
reduced to a composition of $SO(2,2)$ cuts. In this section we will
show how to use this global information in the context of the 
Minkowskian four-dimensional space-time $X^A$.

Let us first consider the case of a particle, where the geodetic
motion is parameterized by (\ref{a22}). In the case $m=1$, we can solve 
the constraints (\ref{a23}) by choosing the following parameterization
\begin{eqnarray}
c_0^A & = & \frac{1}{\sqrt{\Lambda}}
( ch \lambda_1 cos \alpha , - ch \lambda_1 sin \alpha, 
- sh \lambda_1 cos \beta ,  sh \lambda_1 sin \beta ) \nonumber \\
c_1^A & = & \frac{1}{\sqrt{\Lambda}}
( ch \lambda_2 sin \alpha , ch \lambda_2 cos \alpha, 
- sh \lambda_2 sin \beta , - sh \lambda_2 cos \beta ) . 
\label{b1} \end{eqnarray}

Then, we can recast the equations of motion (\ref{a22}) in the
following form where the time evolution has been eliminated, by using
the parameterization (\ref{b1}):

\begin{eqnarray}
X^z e^{i\beta} + \overline{X}^z e^{-i \beta} +  th \lambda_1 
( X^t e^{i \alpha} + \overline{X}^t e^{-i \alpha} ) = 0  \nonumber \\
X^z e^{i\beta} - \overline{X}^z e^{-i \beta} +  th \lambda_2 
( X^t e^{i \alpha} -\overline{X}^t e^{-i \alpha} ) = 0 .
\label{b2} \end{eqnarray}
These equations generalize the well known free motion on the plane, 
characterized by the equation $Z= V T$.

When a source has mass, its geodetic
motion, described by (\ref{b2}), is followed by a generalized deficit
angle or conical cut which can be obtained by looking at the static
cone, defined in the $X^A_0$-coordinates by the cut $ X^z_0 \ 
\rightarrow e^{-2\pi i \mu} \ X^z_0 $, in a different coordinate
system, related by a $SO(2,2)$ linear transformation to the $X^0_A$ one:

\begin{eqnarray}
\widetilde{X}^t_{0+} \ & = & \ ch ( \lambda_1 ) \widetilde{X}^t_{+} 
+ sh ( \lambda_1 ) \widetilde{X}^z_{+} \nonumber \\
\widetilde{X}^t_{0-} \ & = & \ ch ( \lambda_2 ) \widetilde{X}^t_{-} 
+ sh ( \lambda_2 ) \widetilde{X}^z_{-} \nonumber \\
\widetilde{X}^z_{0+} \ & = & \ ch ( \lambda_1 ) \widetilde{X}^z_{+} 
+ sh ( \lambda_1 ) \widetilde{X}^t_{+} \nonumber \\
\widetilde{X}^z_{0-} \ & = & \ ch ( \lambda_2 ) \widetilde{X}^z_{-} 
+ sh ( \lambda_2 ) \widetilde{X}^t_{-} ,
\label{b3} \end{eqnarray}
where we have defined the following combinations:

\begin{eqnarray}
 \widetilde{X}^t_{\pm} & = & X^t e^{i\alpha} \pm \overline{X}^t
e^{-i\alpha} , \ \ \ 
\widetilde{X}^z_{\pm} =  X^z e^{i\beta} \pm \overline{X}^z e^{-i\beta}
\nonumber \\ 
 \widetilde{X}^t_{0\pm} & = & X^t_0 e^{i\alpha_0} \pm \overline{X}^t_0
e^{-i\alpha_0} , \ \ \ 
 \widetilde{X}^z_{0\pm} =  X^z_0 e^{i\beta_0} \pm \overline{X}^z_0 
e^{-i\beta_0} .
\label{b4} \end{eqnarray}

Therefore we obtain as definition of the cut corresponding to the
geodesic motion (\ref{b2}), for arbitrary constants of motion
$ ( \alpha, \beta, \lambda_1, \lambda_2 ) $, the following linear 
transformation

\begin{eqnarray}
\widetilde{X}^t_{+} & \rightarrow & 
( ch^2 \lambda_1 - cos 2\pi\mu \ sh^2 \lambda_1 ) \widetilde{X}^t_{+}
+ sh \lambda_1 ch \lambda_1 ( 1 - cos 2\pi\mu ) 
\widetilde{X}^z_{+} \nonumber \\
& + & i sin 2\pi\mu sh \lambda_1 sh \lambda_2 \widetilde{X}^t_{-} +
i sin 2\pi\mu sh \lambda_1 ch \lambda_2 \widetilde{X}^z_{-} \nonumber \\
\widetilde{X}^t_{-} & \rightarrow & 
( ch^2 \lambda_2 - cos 2\pi\mu \ sh^2 \lambda_2 ) \widetilde{X}^t_{-}
+ sh \lambda_2 ch \lambda_2 ( 1 - cos 2\pi\mu ) 
\widetilde{X}^z_{-} \nonumber \\
& + & i sin 2\pi\mu \ ch \lambda_1 sh \lambda_2 \widetilde{X}^z_{+} +
i sin 2\pi\mu \ sh \lambda_1 sh \lambda_2 \widetilde{X}^t_{+} \nonumber \\
\widetilde{X}^z_{+} & \rightarrow & 
( ch^2 \lambda_1  cos 2\pi\mu - sh^2 \lambda_1 ) \widetilde{X}^z_{+}
+ sh \lambda_1 ch \lambda_1 ( cos 2\pi\mu - 1 ) 
\widetilde{X}^t_{+} \nonumber \\
& - & i sin 2\pi\mu \ ch \lambda_1 ch \lambda_2 \widetilde{X}^z_{-} -
i sin 2\pi\mu \ ch \lambda_1 sh \lambda_2 \widetilde{X}^t_{-} \nonumber \\
\widetilde{X}^z_{-} & \rightarrow & 
( ch^2 \lambda_2 cos 2\pi\mu - sh^2 \lambda_2 ) \widetilde{X}^z_{-}
+ sh \lambda_2 ch \lambda_2 ( cos 2\pi\mu - 1 ) 
\widetilde{X}^t_{-} \nonumber \\
& - & i sin 2\pi\mu ch \lambda_1 ch \lambda_2 \widetilde{X}^z_{+} -
i sin 2\pi\mu sh \lambda_1 ch \lambda_2 \widetilde{X}^t_{+} . 
\label{b5} \end{eqnarray}

A more convenient way to write these monodromy conditions is to apply
the bispinorial formalism, adapted to the group $ SO(2,2) \sim SU(1,1)
\otimes SU(1,1)$. It is not difficult to show that the transformation
\beq \left( \begin{array}{cc} X^t & X^z \\ \overline{X}^z &
\overline{X}^t \end{array} \right) \rightarrow 
 \left( \begin{array}{cc} A_1 & B_1 \\ \overline{B}_1 &
\overline{A}_1 \end{array} \right)
\left( \begin{array}{cc} X^t & X^z \\ \overline{X}^z &
\overline{X}^t \end{array} \right)
\left( \begin{array}{cc} A_2 & B_2 \\ \overline{B}_2 &
\overline{A}_2 \end{array} \right)
\label{b6} \eeq
is a transformation of $SO(2,2)$.

For a spinless particle, whose static cut 
\begin{eqnarray}
X^t & \rightarrow & X^t \nonumber \\
X^z & \rightarrow & e^{-2i \pi \mu } X^z
\label{b7} \end{eqnarray}
can be decomposed as $A_1 = \overline{A}_2 = e^{-i\pi \mu}$, in
general the cut can be defined by the invariant conditions
\beq A_1 + \overline{A}_1 = A_2 + \overline{A}_2 = 2 cos \pi \mu , 
\label{b8} \eeq
that can be solved in such a way to reproduce exactly the Lorentz 
transformation (\ref{b5}) applying the law (\ref{b6}):

\begin{eqnarray}
A_1 \ & = & \ cos \pi \mu  - i ch ( \lambda_1 - \lambda_2 ) sin \pi \mu
 \nonumber \\
B_1 \ & = & \ - i e^{- i( \alpha + \beta )} sh ( \lambda_1 - \lambda_2 )
sin \pi \mu \nonumber \\
A_2 \ & = & \ cos \pi \mu  + i ch ( \lambda_1 + \lambda_2 ) sin \pi \mu
 \nonumber \\
B_2 \ & = & \ - i e^{ i( \alpha - \beta )} sh ( \lambda_1 + \lambda_2 )
sin \pi \mu .
\label{b9} \end{eqnarray}

For a spinning particle, whose static cut 
\begin{eqnarray}
X^t & \rightarrow & e^{2i\pi \sqrt{\Lambda} r_{-}} X^t \nonumber \\
X^z & \rightarrow & e^{2i\pi \sqrt{\Lambda} r_{+}} X^z 
\label{b10} \end{eqnarray}
can be decomposed as $A_1 = - e^{i \pi \sqrt{\Lambda} ( r_{+} + r_{-} )}
\ $ and $ \  \overline{A}_2 = 
- e^{i \pi \sqrt{\Lambda} ( r_{+} - r_{-} )} $ , the general cut can be
defined by the invariant conditions
\beq A_1 + \overline{A}_1 = 
2 cos \pi ( 1 - \sqrt{\Lambda} ( r_{+} + r_{-} ) ) \ \ \ \
 A_2 + \overline{A}_2 =
 2 cos \pi ( 1- \sqrt{\Lambda} ( r_{+} - r_{-}) ),
\label{b11} \eeq
that can be solved similarly to eq. (\ref{b9}): 
\begin{eqnarray}
A_1 \ & = & \ cos \pi ( 1 - \sqrt{\Lambda} ( r_{+} + r_{-} ) ) 
+ i ch ( \lambda_1 - \lambda_2 )
sin \pi \sqrt{\Lambda}  ( r_{+} + r_{-} ) \nonumber \\
B_1 \ & = & \ - i e^{- i( \alpha + \beta )} sh ( \lambda_1 - \lambda_2 )
sin \pi \sqrt{\Lambda} ( r_{+} + r_{-} ) \nonumber \\
A_2 \ & = & \ cos \pi ( 1 - \sqrt{\Lambda} ( r_{+} - r_{-} ) )
- i ch ( \lambda_1 + \lambda_2 )
sin \pi \sqrt{\Lambda} ( r_{+} - r_{-} ) \nonumber \\
B_2 \ & = & \ - i e^{ i( \alpha - \beta )} sh ( \lambda_1 + \lambda_2 )
sin \pi \sqrt{\Lambda} ( r_{+} - r_{-} ) .
\label{b12} \end{eqnarray}

In the case of hyperbolic monodromies, the general invariants relations
(\ref{b11}), defining implicitly two masses, must be substituted with
analogous relations, defining instead two rapidities.
For example, in the case of a spinning black hole, whose static cut
\begin{eqnarray}
 X^0 & \pm  & X^2 \rightarrow e^{\pm 2 \pi \sqrt{\Lambda} r_{+}} 
(  X^0 \pm X^2 )
\nonumber \\ 
 X^1 & \pm  & X^3 \rightarrow e^{\mp 2 \pi \sqrt{\Lambda} r_{-}} 
(  X^1 \pm X^3 ) \label{b15} \end{eqnarray}
can be obtained with the position
\begin{eqnarray}
A_1 & = & ch \pi \sqrt{\Lambda} ( r_{+} + r_{-} ) \ \ \ \ A_2 = ch \pi
\sqrt{\Lambda} ( r_{+} - r_{-} ) \nonumber \\
B_1 & = & sh \pi \sqrt{\Lambda} (r_{+} + r_{-}  ) \ \ \ \ B_2 = sh \pi
\sqrt{\Lambda} ( r_{+} - r_{-} ),  \label{b16} \end{eqnarray}
the general cut can be introduced with the condition:
\begin{eqnarray}
A_1 + \overline{A}_1 & = & 2 ch \pi \sqrt{\Lambda} ( r_{+} + r_{-} )
\nonumber \\
A_2 + \overline{A}_2 & = & 2 ch \pi \sqrt{\Lambda} ( r_{+} - r_{-} ) 
\label{b17} \end{eqnarray}
and it is enough to choose the following parameterization:
\begin{eqnarray}
A_1 & = & ch \pi \sqrt{\Lambda} ( r_{+} + r_{-} ) - i 
sh ( \lambda_1 - \lambda_2 ) sh \pi \sqrt{\Lambda} ( r_{+} + r_{-} ) 
\nonumber\\
B_1 & = & e^{-i ( \alpha + \beta )} ch ( \lambda_1 - \lambda_2 ) sh
\pi \sqrt{\Lambda} ( r_{+} + r_{-} ) \nonumber \\
A_2 & = & ch \pi \sqrt{\Lambda} ( r_{+} - r_{-} ) 
+ i sh ( \lambda_1 + \lambda_2 ) sh \pi \sqrt{\Lambda} ( r_{+} - r_{-} ) 
\nonumber \\
B_2 & = & e^{ i ( \alpha - \beta )} ch ( \lambda_1 + \lambda_2 ) sh
\pi \sqrt{\Lambda} ( r_{+} - r_{-} ) .  
\label{b18} \end{eqnarray}

Now we can ask ourself what is the solution of the two-body
problem in global terms. The result of the composition of two
monodromies, in the case of particles, is of course of the type:

\beq \left( \begin{array}{cc} {X'}^t & {X'}^z \\ \overline{X'}^z &
\overline{X'}^t \end{array} \right) =  M_L 
\left( \begin{array}{cc} X^t & X^z \\ \overline{X}^z &
\overline{X}^t \end{array} \right) M_R ,
\label{b19} \eeq
where to the $M_L, M_R$ matrices it is possible to associate the 
corresponding invariant masses \cite{a21}:

\begin{eqnarray}
cos ( \pi {\cal M}_L ) & = & cos (\pi \mu_1 ) cos ( \pi \mu_2 ) - \frac{P_1^L
\cdot P_2^L}{m_1 m_2} sin ( \pi \mu_1 ) sin ( \pi \mu_2 ) \nonumber \\
cos ( \pi {\cal M}_R ) & = & cos (\pi \mu_1 ) cos ( \pi \mu_2 ) - \frac{P_1^R
\cdot P_2^R}{m_1 m_2} sin ( \pi \mu_1 ) sin ( \pi \mu_2 )
\label{b20} \end{eqnarray}
and we have defined the following vectors, constants of motion: 

\begin{eqnarray}
P_1^L & = & m_1 \gamma_1^L ( 1, v^L_1 ) \ \ \ \ \ 
\gamma_1^L = ch ( \lambda_1 - \lambda_2 ) \ \ \
\gamma_1^L v_1^L = e^{-i (\alpha+\beta)} sh ( \lambda_1 - \lambda_2 )
\nonumber \\
P_1^R & = & m_1 \gamma_1^R ( 1, v^R_1 ) \ \ \ \ \ 
\gamma_1^R = ch ( \lambda_1 + \lambda_2 ) \ \ \
\gamma_1^R v_1^R = e^{-i (\alpha-\beta)} sh ( \lambda_1 + \lambda_2 ) .
\label{b21} \end{eqnarray}

For generic values of the constants of motions, the
left-invariant mass ${\cal M}_L$ will be different from the 
right-invariant mass ${\cal M}_R$ and therefore the composed system 
has spin, other than invariant mass. In fact, by comparing the cut of 
two-body with that of a
spinning particle we obtain that the total mass $\mu_{tot}$ and $J$ are
defined by solving the conditions 
$ { ( 1 - \sqrt{\Lambda} ( r_{+} + r_{-} ) ) }^2 = {\cal M}_{L}^2 $ and
$ { ( 1 - \sqrt{\Lambda} ( r_{+} - r_{-} ) ) }^2 = {\cal M}_{R}^2 $.

An analogous remark can in principle be made for the composition of
hyperbolic monodromies, describing the scattering of $BTZ$ black
holes, however we have no control that such a solution exists, free of
some fancy extra singularity, while in the particle case at least in the
limit  $\Lambda \rightarrow 0$  a physically acceptable solution is known.
To approach the question of the solution for the scattering of black
holes we believe that one has to learn how to give an intrinsic, coordinate
independent, meaning to the horizons.

To make these monodromies more explicit, we are going to solve the
constraint $X^A X_A = \frac{1}{\Lambda}$ with a parameterization which
carries spinorial representations of each $SU(1,1)$:

\beq f \rightarrow \frac{ A_1 f + B_1 }{ \overline{B}_1 f +
\overline{A}_1 } \ \ \ \ \ \ \ g \rightarrow 
\frac{ \overline{A}_2 g + B_2 }{ \overline{B}_2 g + A_2 } . \label{b22} \eeq

Let us choose the following parameterization:

\beq \left( \begin{array}{cc} X^t & X^z \\ {\overline{X}}^z & {\overline{X}}^t 
\end{array} \right) \
= \ h \left( \begin{array}{c} f  \\ 1 \end{array} \right) 
( \overline{g} \ \ \ \ \ 1  )  +
\overline{h}  \left( \begin{array}{c} 1  \\ \overline{f} 
\end{array} \right) ( 1 \ \ \ \  g ) . \label{b23} \eeq
The condition of the constraint (\ref{a0}) implies that
\beq |h|^2 ( 1- f \overline{f} ) ( 1 - g \overline{g} ) = \frac{1}{\Lambda} 
\label{b24} \eeq
The condition of representing the monodromies implies that:

\beq h \sqrt{ \partial_z f \partial_{\overline z} {\overline g} } = H (
z, \overline{z}, t ) , \label{b25} \eeq
where $H$ is a field, invariant under $SU(1,1) \otimes SU(1,1)$, which
can be chosen as
\beq H = \frac{e^{-i T}}{\sqrt{\Lambda}} 
\frac{ {( \partial_z f \partial_{\overline z}
\overline f \partial_z g \partial_{\overline z} \overline g 
)}^{\frac{1}{4}} }{ 
\sqrt{( 1 - f \overline{f} ) ( 1 - g \overline{g} )} } 
\label{b26} \eeq
where we have added an extra phase, the Minkowskian time $T$, which
can play the role of introducing an explicit time evolution of the 
$X^A$-mapping.

We end up with the following solution
\begin{eqnarray}
 \left( \begin{array}{cc} X^t & X^z \\ {\overline{X}}^z & {\overline{X}}^t 
\end{array} \right) & = &  \frac{1}{\sqrt{\Lambda}}
{\left( \frac{\partial_z g \partial_{\overline{z}} \overline{f}}
{\partial_z f \partial_{\overline{z}} \overline{g}} \right)}^{\frac{1}{4}}
\frac{e^{-i T} }{\sqrt{(1- f \overline{f})(1-g\overline{g})}} 
\left( \begin{array}{cc} f \overline{g} & f \\ \overline{g} & 1
\end{array} \right) + \nonumber \\
& +  & \frac{1}{\sqrt{\Lambda}}
{\left( \frac{\partial_z f \partial_{\overline{z}} \overline{g}}
{\partial_z g \partial_{\overline{z}} \overline{f}}
\right)}^{\frac{1}{4}}
\frac{e^{i T} }{\sqrt{(1- f \overline{f})(1-g\overline{g})}} 
\left( \begin{array}{cc} 1 & g \\ \overline{f} & 
\overline{f} g 
\end{array} \right) .
\label{b27} \end{eqnarray}

The analogous case of the static solution of (2+1)-gravity is obtained
by choosing the constants of motion to be ${\cal M}_L = {\cal M}_R$, 
therefore $f=g=Z$
and in this case the parameterization (\ref{b27}) reduces to 
( in the notation $X^A = ( X^0, X^1, X^z, X^{\overline{z}} )$
\beq
X^A = \frac{1}{\sqrt{\Lambda}}
\left( cos T \ \frac{ 1 + Z \overline{Z}}{1- Z \overline{Z}}, \ sin T, 
\ cos T \frac{ 2 Z}{1- Z \overline{Z}},
\ cos T \frac{ 2 \overline{Z}}{1- Z \overline{Z}} \right) .
\label{b28} \eeq
A similar mapping has already been introduced  in \cite{a10}.

Let us first discuss the parameterization (\ref{b28}) 
in the case of one body. To make contact with the 
non-perturbative $N$-body solution of $(2+1)$-gravity 
\cite{a14}, it is
convenient to introduce the spatial conformal gauge $g_{zz}=0$,
which is obtained by (\ref{a3}) with a radial mapping 
\beq r \rightarrow \frac{(1-\mu) r^{(1-\mu)}}{1- \frac{\Lambda}{4} 
r^{2(1-\mu)} } \label{b29} \eeq

Then the $X^A$-mapping (\ref{a3}) becomes

\begin{eqnarray}
X^t \ & = & \ \frac{1}{\sqrt{\Lambda}} \frac{ 1 + \frac{\Lambda}{4} 
r^{2(1-\mu)} }{ 1- \frac{\Lambda}{4} 
r^{2(1-\mu)}} e^{i\sqrt{\Lambda} (1-\mu) t} \nonumber \\
X^z \ & = & \ \frac{z^{1-\mu} }{1- \frac{\Lambda}{4}  r^{2(1-\mu)}} , 
\label{b30} \end{eqnarray}
and in the variables $T$, $Z$ can be rewritten as :

\begin{eqnarray}
sin T & = & a \frac{ 1 + 
\frac{\Lambda}{4} r^{2(1-\mu)} }{ 1- \frac{\Lambda}{4}  r^{2(1-\mu)}} 
\ \ \ \ \ \  a = sin ( \sqrt{\Lambda} (1-\mu) t ) 
\nonumber \\
Z & = & \frac{\sqrt{\Lambda}}{4} \sqrt{ 1- a^2 } z^{1-\mu} \left[
1 + \frac{4}{\Lambda} r^{-2(1-\mu)} \left( 1 - 
\sqrt{ 1 - \frac{\Lambda}{2} 
\left( \frac{ 1+ a^2 }{ 1- a^2 } \right) r^{2(1-\mu)} 
+  \frac{\Lambda^2}{16} r^{4(1-\mu)} } \right)  \right] . \nonumber \\
& & \label{b31} \end{eqnarray}

So we can understand how it is difficult to extend such a mapping
in the $N$-body problem. The polydromic part $z^{(1-\mu)}$ is only a small
piece of the complete solution. Instead in the case of the N-body
solution of $(2+1)$ gravity \cite{a14}-\cite{a15}, the polydromic part
is enough to solve
completely the fields. This involution of the solution is also to be
expected  because there is no $N$-body static solution. The
introduction of an explicit scale parameter allows a static Newtonian
force between point sources. We will show afterwards that a pure
analytic solution to the mapping problem $X^A$ (\ref{b28}) is not
physically consistent.

Moreover there are other problems. In the conformal gauge, there is
a physical limit in which the spatial slice of the universe ends, which is 
visible in the $X^A$ coordinates, since the values of some
$X^A$-coordinates diverges there, but it is not related to the apparent
singularity of the parameterization (\ref{b28}) at the particular
value $ |Z| = 1$, because $cos T$ vanishes also. 
Therefore apart from the incidental
physical limit on the values of the spatial coordinates, depending on
the gauge choice, there is another artificial limit on the spatial 
coordinates, because the parameterization (\ref{b28}) 
cannot globally extended, but it has to substituted with another one
outside its domain of validity.

For more than one body, we must solve the monodromy conditions

\beq Z \rightarrow \frac{a_1 Z + b_1}{\overline{b}_1 Z +
\overline{a}_1} \ \ \ \ \ \ . . . . \ \ \ \  
Z \rightarrow \frac{a_i Z + b_i}{\overline{b}_i Z +
\overline{a}_i} \label{b32} \eeq
that maintain the constraint $|Z|=1$, which defines a well defined closed 
line in the plane. However this line is purely artificial, since the solution
can be continued across it towards the true edge of the spatial
slice of the universe.

As in the one-body case, we are convinced that it is not useful to 
solve these monodromy conditions analytically. Let us suppose to
define a gauge choice such that $Z= Z(z, \xi_i(t))$ has cuts as
defined in eq. (\ref{b32}).  Then the geodesics equations of motions 
for the particles  $\xi_i(t)$, which imply that the values of
the $Z$-mapping coincide at the particle sites with the fixed points
of the $Z$-monodromy, are, in the case of an analytic solution, 
automatically satisfied for an arbitrary motion $\xi_i (t)$ and one is 
tempted to conclude that the dynamics is completely arbitrary against any 
physical intuition and the correspondence with the geodesic limit. 

In the case of $(2+1)$ gravity, the dynamics was defined not only by
the equations of motion but also by the choice of the boundary
conditions of the fields at infinity. In the case of
the cosmological constant, the choice of the boundary conditions 
is a harder problem because the fields do not vanish at infinity.  
Our proposal is to choose a physical gauge, like the one introduced in 
\cite{a14}, and require that in the limit 
$\Lambda \rightarrow 0 $ one recovers
the $N$-body solution of $(2+1)$-gravity, which automatically imposes 
some boundary conditions. Then one can consistently check that the spatial
field equation for $Z= Z(z,\overline{z},t)$ is explicitly
time-dependent, and that this property is enough to produce non-trivial
solutions to the geodesic equations. Detailed analysis will be
presented in a future work.

{\bf Acknowledgements}

It is a pleasure to thank A. Bellini, A. Cappelli and M. Ciafaloni for 
enlightening discussions.

\appendix
\section{Appendix}

 In this appendix we show the equivalence between our method of
solving the geodesics equation and the traditional method of integrate
them. In particular the main point is to rearrange the first
order constants of motion $c^A_i$ in terms of the angular momentum $L$
and the energy $E$, by using the following relations:

\begin{eqnarray}
\sqrt{\Lambda} ( X^0 & \pm & X^2 ) = 
\frac{k_0^{\pm 1}}{\sqrt{r^2_{+}- r^2_{-}}} \
\left\{ 
\begin{array}{ll} 
\left[ \sqrt{\frac{C}{2} \pm G_{-}} 
\ \ cos ( \tau - \tau_0 ) + \sqrt{\frac{C}{2} \mp G_{-}} 
\ \ sin ( \tau - \tau_0 ) \right] & \mbox{if \ $m = 1$} \\
\pm \frac{1}{\sqrt{F_{-}}} \ ( G_{-} \pm F_{-} ( \tau - \tau_0 )) 
& \mbox{if \ $m = 0$, $A \neq 0$ } \\
\mp \left[ \sqrt{\frac{F_{-}-C}{2}} \ ch ( \tau-\tau_0 ) \pm 
\sqrt{\frac{F_{-}+C}{2}} \ sh ( \tau-\tau_0 )\right]
& \mbox{if \ $m = -1$} 
\end{array} \right. \nonumber \\
\sqrt{\Lambda}
( X^1 & \pm & X^3 ) = \frac{k_1^{\pm 1}}{\sqrt{r^2_{+}- r^2_{-}}} \
\left\{ 
\begin{array}{ll}
\pm  \left[ \sqrt{\frac{C}{2} \mp G_{+}} 
\ \ cos ( \tau - \tau_0 ) + \sqrt{\frac{C}{2} \pm G_{+} } 
\ \ sin ( \tau - \tau_0 ) \right] \ \ \ & \mbox{if \ $m = 1$} \\
\frac{1}{\sqrt{F_{+}}} \ ( G_{+} \mp F_{+} ( \tau - \tau_0 ))
& \mbox{if \ $m = 0$, $A \neq 0$ } \\
\left[ \sqrt{ \frac{F_{+} - C}{2}} ch ( \tau - \tau_0 ) \mp
\sqrt{ \frac{F_{+} + C}{2}} sh ( \tau - \tau_0 ) \right]
& \mbox{if \ $m = -1$} 
\end{array} \right. \nonumber \\
k_0 & = & \sqrt{r^2_{+} - r^2_{-}} 
\left\{ 
\begin{array}{ll}
\frac{\sqrt{\Lambda}}{G_{-}} 
\left[ \sqrt{ \frac{C}{2} + G_{-} } \ \ 
( c^0_0 cos \tau_0 + c^0_1 sin \tau_0 ) 
- \sqrt{ \frac{C}{2} - G_{-} } \ \ 
( c^0_1 cos \tau_0 - c^0_0 sin \tau_0 )
\right] 
& \mbox{if \ $m = 1$} \\
\frac{\sqrt{\Lambda}}{\sqrt{ F_{-}}} ( c^0_1 + c^2_1 ) 
& \mbox{if \ $m = 0$, $A \neq 0$ } \\
- \frac{\sqrt{\Lambda}}{G_{-}}
\left[ \sqrt{\frac{ F_{-} + C }{2}} 
( c^0_0 ch \tau_0 + c^0_1 sh \tau_0 )
+ \sqrt{\frac{ F_{-} - C }{2}} 
( c^0_1 ch \tau_0 + c^0_0 sh \tau_0 )
\right]
& \mbox{if \ $m = -1$} 
\end{array} \right. \nonumber \\
k_1 & = & \sqrt{r^2_{+} - r^2_{-}}
\left\{ 
\begin{array}{ll}
\frac{\sqrt{\Lambda}}{G_{+}} 
\left[ \sqrt{ \frac{C}{2} + G_{+} } \ \ ( c^1_1 cos
\tau_0 - c^1_0 sin \tau_0 ) - \sqrt{ \frac{C}{2} - G_{+} } \ \ ( c^1_0 cos
\tau_0 + c^1_1 sin \tau_0 ) \right]
& \mbox{if \ $m = 1$} \\
- \frac{\sqrt{\Lambda}}{\sqrt{F_{+}}} (c^1_1+c^3_1) 
& \mbox{if \ $m = 0$, $A \neq 0$ } \\
\frac{\sqrt{\Lambda}}{G_{+}} 
\left[ \sqrt{ \frac{F_{+} + C}{2} } ( c^1_0 ch \tau_0
+ c^1_1 sh \tau_0 ) - \sqrt{ \frac{F_{+} - C}{2} } ( c^1_1 ch \tau_0
+ c^1_0 sh \tau_0 ) \right] 
& \mbox{if \ $m = -1$} 
\end{array} \right. . \nonumber \\
& & \label{c1} \end{eqnarray}
The definition of $\tau_0$, depending on $m$, is implicit in the
matching with eq. (\ref{a27}). In the derivation of eq. (\ref{c1}) we have
used the fact that these formulas must be compatible with equations 
(\ref{a26}) and (\ref{a27}).

\end{document}